\documentclass[fleqn,twoside]{article}
\usepackage{espcrc2}

\usepackage{graphicx}
\usepackage[figuresright]{rotating}

\newcommand{\AmS}{{\protect\the\textfont2  A\kern-.1667em\lower.5ex\hbox{M}\kern-.125emS}}

\def\lsim{\,\lower2truept\hbox{${< \atop\hbox{\raise4truept\hbox{$\sim$}}}$}\,}
\def\gsim{\,\lower2truept\hbox{${> \atop\hbox{\raise4truept\hbox{$\sim$}}}$}\,}

\title{Sunyaev-Zeldovich effects, free-free 
emission, and imprints on the \\ cosmic microwave background}

\author{C. Burigana\address{IASF/CNR-INAF, via Gobetti 101,
I-40129 Bologna, Italy, {\it burigana@bo.iasf.cnr.it}},
        G. De Zotti\address{OAP/INAF, vicolo dell'Osservatorio 5, 
I-35122 Padova, Italy, {\it dezotti@pd.astro.it}}, 
L. Feretti\address[MCSD]{IRA/CNR-INAF, via Gobetti 101, 
I-40129 Bologna, Italy, {\it lferetti@ira.cnr.it}}}
       
\begin{document}

\begin{abstract}
The signatures from Sunyaev-Zeldovich effects and free-free emission
in the intergalactic and intracluster medium and at
galactic scales probe the structure evolution at various 
cosmic times. The detection of these sources and, possibly, the  
precise imaging of their spatial structure at the high resolution 
and sensitivity achievable with SKA will greatly contribute 
to the comprehension of crucial cosmological and astrophysical aspects, 
as the physical conditions of early ionized halos, quasars
and proto-galactic gas and the mapping of the thermal and density 
structure of clusters of galaxies, with an extremely 
accurate control of extragalactic radio sources,
which represents also a useful by-product for future CMB spectrum 
experiments devoted to probe the thermal plasma history at early 
times.
\vspace{1pc}
\end{abstract}

\maketitle

\section{INTRODUCTION}
\label{intro}

The thermal plasma in the intergalactic and intracluster medium and at
galactic scales leaves imprints on the cosmic microwave background (CMB) 
through the Thomson scattering
of CMB photons on  hot electrons and the free-free emission.
Although not specifically devoted to CMB studies, because of its 
high resolution and the limited high frequency coverage, 
the extreme sensitivity and resolution of SKA may be fruitfully 
used for a detailed mapping of the above effects on dedicated sky areas.
A particularly 
accurate control of the extragalactic radio sources
represents a useful by-product for future CMB spectrum 
experiments devoted to probe the thermal plasma history at high redshifts. 
The SKA highest frequencies ($\simeq 20$~GHz) are, of course, the 
most advantageous to detect Sunyaev-Zeldovich effects on the CMB because 
of the steeper decrease of the synchrotron radio emission with 
the frequency.

\section{THERMAL AND KINETIC SUNYAEV-ZELDOVICH EFFECT TOWARDS 
GALAXY CLUSTERS }
\label{sz_clust}

The scattering of CMB photons from hot electrons in galaxies 
and clusters of galaxies produces a frequency dependent change
in the CMB brightness. If the hot electron gas is globally  
at rest with respect to the observer only the thermal 
Sunyaev-Zeldovich (SZ) effect \cite{sz72} (see also \cite{rephaeli95}) 
will be present; 
differently, a bulk 
peculiar motion, $V_r$, of the hot electron gas produces  a kinetic
SZ effect. In the Rayleigh-Jeans (RJ) region the first effect 
produces a decrement of the surface brightness,
$\Delta I_{th}$, towards the cluster. The second effect
produces either a decrement or an increment, $\Delta I_k$,
depending on the direction of the cluster velocity with respect to 
the observer.
Neglecting relativistic corrections: 

\begin{equation}
\Delta I_{th} = I_0 y g(x)
\end{equation}
and   
\begin{equation}
\Delta I_k = -I_0 (V_r/c) \tau_e h(x) \, ,
\end{equation}
where   
$I_0=(2h_P/c^2)(k_B T_{CMB}/h)^3$. Here 
\begin{equation}
\tau_e=\int n_e \sigma_T dl
\end{equation}
and 
\begin{equation}
y=\int (k_BT_e/m_ec^2) n_e \sigma_T dl
\end{equation}
are respectively the Thomson optical depth and the Comptonization 
parameter \cite{ZeldovichSunyaev1969} 
integrated over the cluster along the line of sight, $n_e$ being the 
electron density, and
\begin{equation}
h(x)=x^4e^x/(e^x-1)^2
\end{equation}
\begin{equation}
g(x)=h(x) [x(e^x+1)/(e^x-1)-4] \, ,
\end{equation}
where $x=h_P\nu/k_BT_{CMB}$ is a dimensionless photon frequency.
The two effects have a different frequency dependence 
that in principle allows their separation through 
multi-frequency observations. 
SKA observations in the RJ regime (where $h(x) \sim g(x) 
\rightarrow x^2$) 
can be combined with millimetric observations 
(in particular $g(x) \simeq 0$ and $h(x)$ is maximum 
at $\sim 217$~GHz).

Many thousands of galaxy clusters
will be identified by XMM ($\sim 10^3$), 
Planck ($\sim 10^4$), and SDSS ($\sim 5 \, 10^5$). The typical angular 
sizes of galaxy clusters range from $\sim$~arcmin to
few tens of arcmin.

The SKA sensitivity and resolution mainly depends on the used array 
collecting area.
By considering a frequency band of $\simeq 4$~GHz at 20~GHz,
the whole instrument collecting area 
will allow to reach a (rms) sensitivity of $\simeq 40$~nJy 
in one hour of integration with 
an angular resolution of $\simeq 1$~mas
(considering $\simeq 3000$~km maximum baseline).
By using only about 50\% of the collecting area
within $\simeq 5$~km, the (rms) sensitivity
in one hour of integration 
is $\simeq 80$~nJy with a  resolution of $\simeq 0.6''$.

Although the major role on the study of the SZ effects
towards galaxy clusters will be played by 
dedicated telescopes operating at $\simeq$ arcmin resolutions with 
frequency coverages up to $\simeq$ millimetric wavelengths
\cite{jones}, 
with the 50\% of the SKA collecting area it will be possible 
to accurately map the 
SZ effect  \cite{subra_ekers}
of each considered cluster, particularly at moderately high redshifts, 
with an extremely precise subtraction of discrete radio sources
(see also Sect.~\ref{cmb_spec}).
The combination with X-ray images, 
in particular with those expected by   
the wide field imager (WFI) on board the 
X-ray Evolving Universe Spectroscopy (XEUS) 
satellite~\footnote{http://sci.esa.int/science-e/www/area/index.cfm?fareaid=25}
by ESA designed to reach  
a resolution of $0.25''$ on a FOV of $5'-10'$,
will allow to accurately map the thermal and density 
structure of the gas in galaxy clusters. 

Finally, we note that it is also possible to study 
the SZ effect (both thermal and kinetic) from clusters in a 
statistical sense, namely through its contribution to the 
angular power spectrum, $C_\ell$, (see Appendix A) of the CMB 
secondary anisotropies.  
This topic has been investigated in several papers 
(see, e.g., 
\cite{ostrikervishniac,vishniac,gnedin2000,springel2001,dasilva2001,ma_fry_2002}). 
At sub-arcmin scales (i.e. at multipoles $\ell \gsim 10^4$) 
secondary anisotropies from thermal 
(more important at  $\ell \lsim {\rm few} \times 10^4$)
and kinetic 
(more important at  $\ell \gsim {\rm few} \times 10^4$)
SZ effect dominate over CMB primary anisotropy whose
 power significantly 
decreases at multipoles $\ell \gsim 10^3$
because of photon diffusion (Silk damping effect \cite{silk68}).
Their angular power spectrum 
at $\ell \sim 10^4-10^5$ ($\approx 10^{-12} - 10^{-13}$ in terms
of dimensionless $C_\ell \ell (2 \ell+1)/4\pi$)
could be in principle investigated  
with the sensitivity achievable with SKA (see Fig. 6 and
Appendix A).  
On the other hand, at the SKA resolution and sensitivity 
the contribution to fluctuations from foreground sources 
(both diffuse radio emission, SZ effects, and free-free
emitters) at galaxy scales probably dominates over the SZ effect from 
clusters.

\section{THERMAL SUNYAEV-ZELDOVICH EFFECT AT GALAXY SCALE}
\label{sz_gal}

The proto-galactic gas
is expected to have a large thermal energy content, leading to a
detectable SZ signal, both when the protogalaxy collapses with the
gas shock-heated to the virial temperature 
\cite{ReesOstriker1977,WhiteRees1978}, and in a later phase as
the result of strong feedback from a flaring active nucleus
(see, e.g.,  \cite{Ikeuchi1981,Natarajanetal1998,NatarajanSigurdssson1999,Aghanimetal2000,Plataniaetal2002,Lapietal2003}. 
The astrophysical 
implications of these scenarios have been recently investigated 
by \cite{dezottiSZ}.

A fully ionized gas  with a thermal energy density
$\epsilon_{\rm gas}$ within the virial radius
\begin{eqnarray} R_{\rm vir} &=& \left({3 M_{\rm vir}\over 4\pi
\rho_{\rm vir}}\right)^{1/3}  \\
& \simeq & 1.6\, 10^2 h^{-2/3} (1+z_{\rm vir})^{-1} \nonumber \\
&& 
\left({M_{\rm vir}\over 10^{12} M_\odot}\right)^{1/3}\
\hbox{kpc} \nonumber \, ,
\end{eqnarray}
transfers to the CMB an amount 
$\Delta \epsilon \simeq  (\epsilon_{\rm gas} / t_c)   2 (R_{\rm vir}/c)$ 
%
of thermal energy through
Thomson scattering producing a Comptonization parameter 
\cite{ZeldovichSunyaev1969} 
$y \simeq (1/4) \Delta \epsilon / \epsilon_{\rm CMB}$.
%
Here $h$ is the Hubble constant in units of
$100~\hbox{km}\,\hbox{s}^{-1}\,\hbox{Mpc}^{-1}$ and $\rho_{\rm
vir}\simeq 200 \rho_u$, $\rho_u=1.88 \times 10^{-29}h^2(1+z_{\rm
vir})^3\,\hbox{g}\,\hbox{cm}^{-3}$ is the mean density of the
universe at the virialization redshift $z_{\rm vir}$;
$\epsilon_{\rm CMB} = a_{BB} T_{\rm CMB}^4 \simeq 4.2 \times
10^{-13}(1+z)^4\,\hbox{erg}\,\hbox{cm}^{-3}$, $a_{BB}$ being the
black-body constant and $T_{\rm CMB}= T_0(1+z)$ the temperature of
the CMB;
$t_c$ is the gas cooling time by Thomson scattering.

Assuming the binding energy ($E_{\rm b, gas}= M_{\rm
gas} v_{\rm vir}^2$, $v_{\rm vir}=162 h^{1/3}(1+z)^{1/2}(M_{\rm
vir}/10^{12}M_\odot)^{1/3}$ km~s$^{-1}$ being the
circular velocity of the galaxy at its virial radius 
\cite{Navarroetal1997,Bullocketal2001})
to characterize the thermal energy content of the
gas, $E_{\rm gas}$,
the amplitude of
the SZ dip in the RJ region can be written as:
\begin{eqnarray}
\left|\Delta T\right|_{\rm RJ} &=& 2yT_{\rm CMB} \\
 & \simeq &  1.7
\left({h\over 0.5}\right)^{2} \left({1+z_{\rm vir}\over 3.5}\right)^3 
\nonumber \\
&& {M_{\rm gas}/M_{\rm vir}\over 0.1} {M_{\rm vir}\over
10^{12} M_\odot } {E_{\rm gas} \over E_{\rm b, gas}}\
\mu\hbox{K} \nonumber \, .
\label{DeltaT}
\end{eqnarray}

This SZ effect shows up on small (typically
sub-arcmin) angular scales. 

Quasar-driven blast-waves could inject into the ISM an amount of
energy several times higher than the gas binding energy, thus
producing larger, if much rarer, SZ signals.
A black-hole (BH) accreting a mass $M_{\rm BH}$ with a mass to
radiation conversion efficiency $\epsilon_{\rm BH}$ releases an
energy $E_{\rm BH}=\epsilon_{\rm BH}M_{\rm BH}c^2$. We adopt the
standard value for the efficiency $\epsilon_{\rm BH}=0.1$ and
assume that a fraction $f_h=0.1$ of the energy is fed in kinetic
form and generates strong shocks turning it into heat. 

Using the recent re-assessment by \cite{Tremaineetal2002} of
the well known correlation between the BH mass and the stellar
velocity dispersion,
%
$M_{\rm BH} = 1.4 \times 10^8\,\left( {\sigma / 200\,{\rm 
km/s}}\right)^{4}\ 
{\rm M}_\odot$~,
%
we get
\begin{eqnarray}
{E_{\rm BH}\over E_{\rm b, gas}} & \simeq & 4.7  \left({h\over
0.5}\right)^{-2/3}{\epsilon_{\rm BH}\over 0.1}\, {f_h\over 0.1} \,
{1+z_{\rm vir}\over 3.5}\,  \\
&& \left({M_{\rm gas}/M_{\rm vir}\over
0.1}\right)^{-1} \left({M_{\rm vir}\over 10^{12}
M_\odot}\right)^{-1/3} \nonumber \, .  
\label{ratioT}
\end{eqnarray}
The amplitude of the SZ dip in the RJ region due to
quasar heating of the gas is then estimated as:
\begin{eqnarray}
&& \left|\left({\Delta T \over T}\right)_{\rm RJ}\right|  \simeq 
1.8\times 10^{-5} {f_h \over 0.1}  \\
&& \left({h\over 0.5}\right)^2
\left({\epsilon_{\rm BH} \over 0.1}\right)^{1/2} \left({E_{\rm
BH}\over 10^{62}}\right)^{1/2}\left({1+z \over 3.5}\right)^{3} \nonumber \, . 
\label{DeltaTqso}
\end{eqnarray}
Following \cite{Plataniaetal2002}, we adopt an isothermal
density profile of the galaxy. The virial radius, encompassing a
mean density of $200\rho_u$, is then:
\begin{eqnarray}
R_{\rm vir} & \simeq & 120 \left({h\over 0.5}\right)^{-1}
\left({E_{\rm BH}\over 10^{62}}\right)^{1/4}  \\
&& \left({\epsilon_{\rm
BH}\over 0.1}\right)^{-1/4}\left(1+z_{\rm vir} \over
3.5\right)^{-3/2}\ \hbox{kpc} \nonumber \, , 
\label{Rg}
\end{eqnarray}
corresponding to an angular radius:
\begin{eqnarray}
\theta_{SZ} & \simeq & 17'' \left({E_{\rm BH}\over
10^{62}}\right)^{1/4} \left({\epsilon_{\rm BH}\over
0.1}\right)^{-1/4}  \\
&& \left(1+z_{\rm vir} \over
3.5\right)^{-3/2}{d_A(2.5)\over d_A(z)} \nonumber \, , 
\label{theta}
\end{eqnarray}
where $d_A(z)$ is the angular diameter distance.
                                                                                
The angular scales of these SZ signals from galaxies are of the order
of $\approx$ 10$^{\prime\prime}$, then of 
particular interest for a detailed mapping with the SKA and XEUS 
in the radio and X-ray,
respectively.
The probability of observing these SZ sources 
on a given sky field 
at a certain flux detection level  and the
corresponding fluctuations are mainly determined by 
the redshift dependent source number density 
$\phi_{\rm SZ}(S_{\rm SZ},z)$
per unit interval of the SZ 
(decrement) flux $S_{\rm SZ}$.
The lifetime of the considered SZ sources 
is crucial to determine their number density. 

For quasar-driven blast-waves the lifetime of the active phase,
$t_{\rm SZ}$, is approximately equal to the time for the shock
to reach the outer boundary of the host galaxy.
Assuming a self-similar blast-wave expanding in a medium with an
isothermal density profile, $\rho \propto r^{-2}$, we have:
\begin{eqnarray}
t_{\rm SZ} & \simeq & 1.5\times 10^8 \left({h\over 0.5}\right)^{-3/2}
\left({E_{\rm BH} \over
10^{62}\hbox{erg}}\right)^{1/8}  \\
&& \left({\epsilon_{\rm BH}\over
0.1}\right)^{-5/8}\left({f_h\over
0.1}\right)^{-1/2}\left({1+z\over 3.5}\right)^{-9/4}\ \hbox{yr} 
\nonumber \, .
\label{tSZ}
\end{eqnarray}

The evolving B-band luminosity function of quasars, $\phi(L_B,z)$, 
can be then adopted to derive the 
source number density $\phi_{\rm SZ}(S_{\rm SZ},z)$ according to
\begin{equation}
\phi_{\rm SZ}(S_{\rm SZ},z) = \phi(L_B,z) {t_{\rm SZ} \over t_{\rm
q,opt}}\,{d L_B \over d S_{\rm SZ}} \ , \label{phiSZ}
\end{equation}
where $L_B(S_{\rm SZ},z)$ is the blue luminosity of a quasar at
redshift $z$ causing a (negative) SZ flux $S_{\rm SZ}$, 
and $t_{\rm q,opt}$
is the duration of the optically bright phase of the quasar
evolution.

For the proto-galactic gas $t_{\rm SZ}$ should be replaced by the gas 
cooling time, $t_{\rm cool}$.
Assuming that quasars can be used as
effective signposts for massive spheroidal galaxies in their early
evolutionary phases \cite{Granatoetal2001} 
and that they emit at the Eddington limit
and using 
the relation by \cite{Ferrarese2002} between 
the mass of the dark-matter halo, $M_{\rm vir}$, and the 
mass of the central black-hole,
${M_{\rm BH}/ 10^8 M_\odot} \sim 0.1 
(M_{\rm vir} / 10^{12} M_\odot)^{1.65}$,
the number density of sources with gas at virial temperature
can be straightforwardly related to the quasar luminosity function
$\phi(L_B,z)$. 

\begin{figure}[htb]
\includegraphics[angle=0.,width=7.cm,height=7.cm]{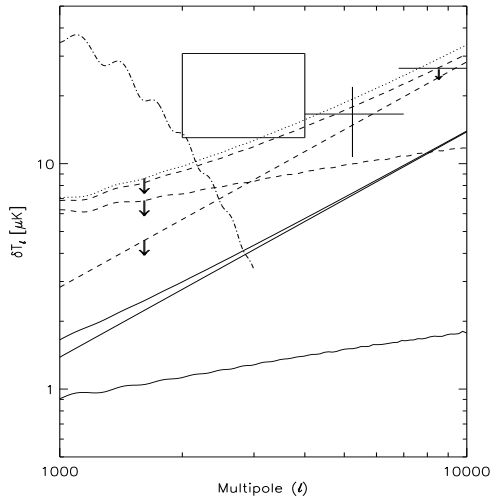}
\caption{Angular power spectrum of SZ effects at 30~GHz
compared to CMB primary fluctuation power spectrum and 
CBI \cite{Masonetal2003} (box)
and BIMA \cite{Dawsonetal2002} (data points)
measures. Solid lines represent 
clustering (bottom line), Poisson (middle line) and global 
(upper line) contributions from quasar driven blast-waves.
Dashed lines represent clustering (bottom
 line at high $\ell$), Poisson (middle line at high $\ell$) 
and global (upper line) contributions from 
proto-galactic gas. The latter are actually upper limits since,
 because of the uncertainty in the cooling time,
the extreme assumption that $t_{\rm cool}=t_{\rm exp}$ has been adopted in 
the computation.
Dots refer to the overall contribution.} 
\label{fig:sz_gal}
\end{figure}

In spite of the many uncertainties of these models, it is remarkable
that the CMB fluctuations
(dominated at small scales by the Poisson contribution) 
induced by the SZ effect of these 
source populations could contribute to the CBI 
anisotropy measure and, in particular, could explain the angular 
power spectrum (see also Appendix A) found by BIMA at multipoles 
$\ell \approx (4-10) \times 10^3$ (see Fig.~\ref{fig:sz_gal}).

\begin{figure}[htb]
\includegraphics[angle=0.,width=7.cm,height=7.cm]{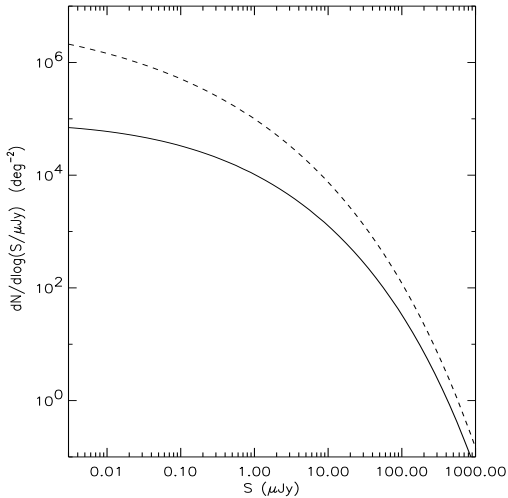}
\caption{Number count predictions at 20~GHz for SZ effects 
as function of the absolute value of the flux 
from proto-galactic gas heated at the virial temperature (dashes)
assuming $M_{\rm gas}/M_{\rm vir} = 0.1$
and from quasar driven blast-waves (solid line).
The exponential model for the evolving luminosity function 
of quasars is derived by
\cite{Pei1995} 
for an optical spectral index of quasars $\alpha=0.5$ ($S_\nu \propto
\nu^\alpha$). The parameters have been set at $\epsilon_{\rm
BH}=0.1$, $f_h=0.1$, $k_{\rm bol}=10$, $t_{\rm q,opt}=10^7\,\hbox{yr}$.}
\label{fig:n_sz_gal}
\end{figure}

A direct probe of these models and, possibly, their accurate knowledge 
through a precise high resolution imaging is then 
of particular interest.
Fig.~\ref{fig:n_sz_gal} shows the number counts 
at 20~GHz predicted by these models:
in a single SKA FOV about few~$\times 10^2 - 10^3$ SZ sources with fluxes
above $\sim 100$~nJy could be then observed in few hours of 
integration.
Given the typical source sizes, 
we expect 
a blend of sources in the SKA FOV 
at these sensitivity levels, while much shorter integration times,
$\sim$~sec, on many FOV would allow to obtain much larger maps 
with a significant smaller number of resolved SZ sources per FOV.
Both surveys on relatively wide sky areas and deep exposures
on limited numbers of FOV are interesting and easily 
obtainable with SKA.

A different scenario to jointly explain the power excess 
found by BIMA and the high redshift reionization detected 
by WMAP \cite{kogutetal03}
($z_{reion} \sim 15-20$) has been proposed by \cite{OH03}. 
It involves hot gas winds powered by pair-instability supernovae (SN)
explosions from the first generation of very massive stars at very 
low metallicity able to photoevaporate the gas in the halo potential.
The SN remnants should then dissipate their energy in the 
intergalactic medium (IGM) and about 30-100\% of their energy  
would be transferred to the CMB via Compton cooling.
The resulting SZ effect from this sources is 
relevant in statistical sense. It is claimed to explain 
the high $\ell$ BIMA excess of the CMB angular power spectrum
and to be able to generate a global Comptonization distortion
parameter $y \sim {\rm few} \times 10^{-6}$. However, it
is estimated to be too faint ($\approx {\rm few} \times 
10^{-2}$~nJy) to be observable even by SKA.

\section{SKA CONTRIBUTION TO FUTURE CMB SPECTRUM EXPERIMENTS}
\label{cmb_spec}

\begin{figure}[htb]
\includegraphics[angle=0.,width=7.cm,height=7.cm]{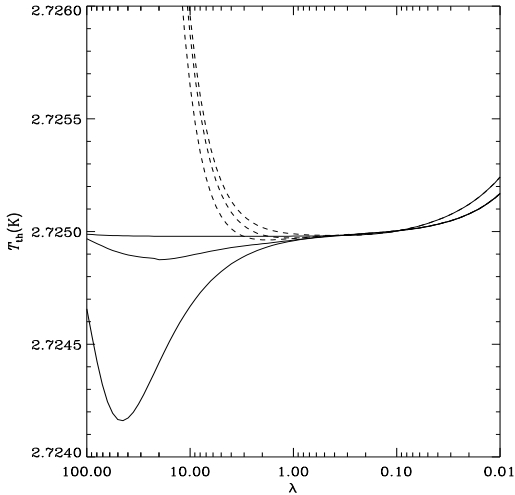}
\caption{CMB distorted spectra as functions of the wavelength
$\lambda$ (in cm) in the presence of
a late energy injection with $\Delta \epsilon / \epsilon_i = 5 \times 
10^{-6}$ plus an early/intermediate
energy injection with $\Delta \epsilon / \epsilon_i = 5 \times 10^{-6}$
occurring at the ``time'' Comptonization 
parameter
$y_h=5, 1, 0.01$ (from the bottom to the top;
in the figure the cases at $y_h=5$ and 1 are indistiguishable
at short wavelengths; solid lines) and plus a
free-free distortion
with  $y_B=10^{-6}$ (dashes).
$y_h$ is defined by Eq.~(4) 
with $dl=cdt$ and $T_e=T_{CMB}$ when the integral
is computed from the time of the energy injection
to the current time.} 
\label{fig:cmb_dist}
\end{figure}

The current limits on CMB spectral distortions and the constraints on 
energy dissipation processes $|\Delta \epsilon / \epsilon_i| \lsim 
10^{-4}$ in the plasma \cite{SB02}
are mainly set by COBE/FIRAS \cite{mather90,fixsen96}.
CMB spectrum experiments from space,
DIMES \cite{KOG96} (see also \cite{KOG03})
at $\lambda \gsim 1$~cm and FIRAS~II \cite{FM02}
at $\lambda \lsim 1$~cm, 
have been proposed with an accuracy potentially able to 
constrain (or probably detect) energy exchanges 10--100 times 
smaller than the FIRAS upper limits.
In particular, experiments like DIMES may probe dissipation
processes at early times ($z \gsim 10^5$)
resulting in Bose-Einstein like distortions 
\cite{SZ70,DD80,BDD91a} 
and free-free 
distortions \cite{bart_stebb_1991}
possibly generated by heating (but, although disfavoured
by WMAP, in principle  
also by cooling \cite{stebb_silk}) 
mechanisms at late epochs ($z \lsim 10^4$),
before or after the recombination era \cite{BS03a}.
The high redshift reionization detected by WMAP \cite{kogutetal03}
supports the existence of late coupled Comptonization 
(with $\Delta \epsilon / \epsilon_i \simeq 4y \approx {\rm few} \times 
10^{-6}$,  $\epsilon_i$ being the CMB energy density before the beginning
of the dissipation process) 
and free-free distortions (with a highly model dependent
amplitude) \cite{buriganaetal04}. 
Typical distorted spectra 
potentially detectable by DIMES
are shown in Fig.~\ref{fig:cmb_dist}.
To firmly observe such small
distortions the Galactic and extragalactic foreground contribution
should be accurately modelled and subtracted. 

Recent progress on radio source counts at 1.4~GHz have been presented
in \cite{prandonietal01}. On the other hand, the very faint tail
of radio source counts is essentially unexplored and their 
contribution to the radio background at very low brightness 
temperature is not accurately known.   
For illustration, by assuming differential source number counts,
$N(S)$, given by ${\rm log} N(S)/\Delta N_0 \sim a {\rm log} S + b$,
with $\Delta N_0 \sim 150 S^{-2.5}$~sr$^{-1}$~Jy$^{-1}$ ($S$ in Jy)
\cite{toffolattietal98},
for $a \sim 0.4-0.6$ and $b \sim - (0.5-1)$, we find a contribution
to the radio background at 5~GHz 
from sources between $\sim 1$~nJy and $\sim 1 \mu$Jy 
between few tens of $\mu$K and few mK.
These signals are clearly negligible compared to the accuracy of current 
CMB spectrum experiments, in particular at $\lambda \gsim 1$~cm,
but are significant at the accuracy level on CMB distortion 
parameters potentially achievable with experiments like DIMES. 
This effect is small compared to the Galactic 
radio emission, whose accurate knowledge currently represents 
the major astrophysical problem in CMB spectrum experiments,
but, differently from Galactic emission, it is 
isotropic at the angular scales of few degrees 
and can not be then subtracted from the CMB monopole 
temperature on the basis of its angular correlation properties. 
With accurate absolute measures on a wide frequency coverage
a fit including both CMB distorted spectra and astrophysical 
contributions can be searched (see \cite{SB02} for an application
to FIRAS data) but  a direct radio background estimate
from precise number counts will certainly improve the robustness
of this kind of analyses.      

The SKA sensitivity at 20~GHz will allow 
the detection (to $5\sigma$) 
of sources down to a flux level of 
$\simeq 200$~nJy
($\simeq 60, 20, 6$~nJy) 
in 1 (10, $10^2$, $10^3$) hour(s) of integration 
over the $\simeq 1$~mas (FWHM) resolution element;
similar numbers (from $\simeq 250$ to 8~nJy 
in an integration time from 1 to $10^3$ hours,
respectively) but on a resolution element about 10 times
larger will be reached at $\approx$~GHz frequencies
by using a frequency bandwidth of about 25\%.

Therefore, the SKA accurate determination of source number 
counts down to very faint fluxes can directly help the solution of
one fundamental problem of the future generation of CMB 
spectrum space experiments at $\lambda \gsim 1$~cm. 

\section{FREE-FREE EMITTERS}
\label{ff_emitt}

The epoch of reionization has been extensively studied 
in the recent years and in particular since the WMAP 1-yr
data release. Reionization affects the CMB 
both in anisotropies at large and small scales and in 
the spectrum.
The understanding of the ionizing emissivity of collapsed objects and 
the degree of gas clumping is crucial for reionization models.
The observation of diffuse gas and Population III objects
in thermal bremstrahhlung as a direct
probe of these quantities has been investigated by 
\cite{OH99}. Free-free emission produces both global 
and localized spectral distortion of the CMB.
A natural way to distinguish between free-free distortion 
by ionized halos rather than by diffuse ionized IGM
is represented by observations at high resolution of dedicated sky areas 
and 
by the fluctuations in the free-free background.
In the model by \cite{OH99}  halos collapse and form
a starburst lasting $t_{o}=10^{7}$ yr, then recombine and no longer
contribute to the free-free background.
By adopting a Press-Schechter model 
\cite{PressSchechter1974,Bondetal1991}
for the number density of collapsed 
halos per mass interval, $d n_{PS}/dM$, \cite{OH99} 
exploited the expression by \cite{sasaki1994}
for the collapse rate of halos per mass interval per unit comoving volume:
\begin{equation}
\frac{d\dot{N}^{form}}{dM}(M,z)=
\frac{1}{D}\frac{dD}{dt}\frac{d n_{PS}}{dM}(M,z)
\frac{\delta_{c}^{2}}{\sigma^{2}(M)D^{2}} \, ; 
\end{equation}
here
$D(z)$ is the growth factor and $\delta_{c}=1.7$ is the
threshold above which mass fluctuations collapse.
The expected comoving number density of ionized halos in a given flux
interval as a function of redshift 
\begin{equation}
\frac{dN_{\rm halo}}{dS dV}(S,z)= \int_{t(z)}^{t(z)-t_{o}} dt
\frac{d \dot{N}^{form}}{dM} \frac{dM}{dS}
\end{equation}
can be then computed given the expected flux from a halo of mass M at 
redshift $z$,  $S=S(M,z)$, and the starburst duration, $t_{o}$.
Adopting a cut-off mass for a halo to be ionized of
$M_{*}=10^{8} (1+z/10)^{-3/2} M_{\odot}$ 
(the critical mass needed to attain a virial temperature 
of $10^{4}$~K to excite atomic hydrogen cooling), 
\cite{OH99} computed the number counts of sources above the flux limit 
$S_{c}$ from the zeroth
moment of the intensity distribution 
moments due to sources
above a redshift $z_{\rm min}$,
\begin{eqnarray}
&& \langle S^{n}(>z_{min},S_{c}) \rangle \\
&&  = \int_{z_{\rm min}}^{\infty} dz
\int_{S_{\rm min}(z)}^{S_{max}} dS
\frac{dN_{\rm halo}}{dSdV} \frac{dV}{dz d\Omega} S^{n} \nonumber \, ,
\label{moments}
\end{eqnarray}
by setting $S_{max} \rightarrow \infty$ 
and $S_{min}(z)={\rm max}(S_{c},S_{*}(z))$, where 
$S_{*}(z)$ denotes the flux from a halo of
minimum mass $M_{*}$ at redshift $z$.

The relation 
$\dot{N}_{\rm recomb}= \alpha_{B} \langle n_{e}^{2} \rangle V \approx
(1-f_{esc})\dot{N}_{\rm ion} \, ,$
between the production rate of
recombination line photons, $\dot{N}_{\rm recomb}$, 
and the production rate of ionizing photons,
$\dot{N}_{\rm ion}$,
(here $\alpha_{B}$ is the 
recombination coefficient 
and $f_{esc}$ ($\approx {\rm some} \%$) is the escape fraction for 
ionizing photons)
implies that
the source luminosities in H$\alpha$ and free-free emission 
($\propto n_{e}^{2}V$)
are proportional to the production rate of ionizing photons.
Over a wide range of nebulosity conditions \cite{HummerStorey1987} 
found that $\simeq$~0.45~H$\alpha$ photons are emitted per Lyman 
continuum photon; thus 
${\rm L(H \alpha)} = 1.4 \times 10^{41}
{\dot{N}_{\rm ion}}/({10^{53} {\rm
ph \, s^{-1}}})  {\rm erg \, s^{-1}}$. 
Given the free-free volume
emissivity
\cite{Rybicki_Lightman_1979}
in the case of an approximate mild temperature 
dependence with a power law (a velocity averaged Gaunt factor
$\bar{g}_{ff}=4.7$ is assumed),
$\epsilon_{\nu} = 3.2 \times 10^{-39} n_{e}^{2} ({T}/{10^{4} K})^{-0.35}
\ {\rm erg \,s^{-1} \,cm^{-3} \, 
Hz^{-1}}$~,
it is found
\begin{equation}
L_{\nu}^{ff}=1.2 \times 10^{27} \frac{\dot{N}_{\rm
ion}}{10^{53} {\rm ph \, s^{-1}}} {\rm erg \, s^{-1} \,
Hz^{-1}} \, .
\label{L_free}
\end{equation}
Assuming the starburst model of \cite{HaimanLoeb1998} 
normalised to the observed metallicity
$ 10^{-3} Z_{\odot} \le Z \le  10^{-2} Z_{\odot}$ of
the IGM at $z \sim 3$ (resulting into 
a constant fraction of the gas mass turning into stars,
$1.7 \% \le f_{star} \le 17\%$,
in a starburst which fades after $\sim 10^{7}$~yr),  \cite{OH99} 
derived  a production rate of ionizing photons as a function of halo mass
given by:
\begin{equation}
\dot{N}_{\rm ion}(M)=2 \times 10^{53} \,
{f_{star} \over 0.17} \, {M \over 10^{9} M_{\odot}} 
\, {\rm ph} \,  {\rm s^{-1}} \, ,
\label{Ndot_scaling}
\end{equation}
which specifies the above free-free ionized halo
luminosity. The corresponding flux is then:
\begin{eqnarray}
&& S_{\rm ff} = \frac{L_{\nu}^{\rm ff}}{4 \pi d_{L}^{2}}
(1+z) \\
& \approx & 2.5 \left( \frac{1+z}{10} \right)^{-1}
\frac{\rm M} {10^{9} {\rm M_{\odot}}}
\left( \frac{\rm T} {10^{4} \, {\rm K}} \right)^{-0.35} {\rm nJy} 
\nonumber \, .
\label{J_free_free}
\end{eqnarray}

\begin{figure}[htb]
\includegraphics[angle=0.,width=7.cm,height=7.cm]{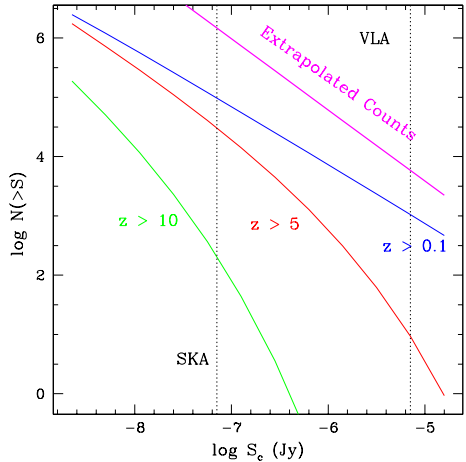}
\caption{Number of sources which may be detected in the $1^{\circ}$
by SKA, as a function of the threshold flux
$S_{c}$. Realistic limiting fluxes for point source detection are shown. 
The extrapolated source counts from \cite{Partridgeetal1997} are also 
shown.
From \cite{OH99}.}
\label{Num_sources_free}
\end{figure}

Clearly, SKA will allow to detect only bright sources
with deep exposures. 
The ionized halo number counts can be calculated from
Eq.~(17).
The result by \cite{OH99} is reported in 
Fig.~\ref{Num_sources_free}:
SKA should be able to detect $\sim 10^{4}$
individual free-free emission sources with $z>5$ 
in 1 square degree above a source detection threshold of 70~nJy. 
The redshift information from the Balmer line emission detectable
by the Next Generation Space Telescope (NGST) can be used to discriminate
ionized halos from other classes of radio sources.

Ionized halos may contribute 
to the temperature fluctuations. In particular, the Poisson contribution
is predicted to be larger (smaller) than the clustering one at  
scales smaller (larger) than $\sim 30''$ \cite{OH99}. On the other hand,
both are likely dominated by the radio source contribution. 

Finally, the integrated emission from ionized halos produces a global CMB 
spectral distortion,
$\Delta T_{ff}=c^{2}\langle S \rangle /2k_B\nu^{2}$, 
that can be computed from the 
mean sky averaged signal $\langle S \rangle$.
By using Eq.~(17)
(since 
no point source removal is feasible at degree scales) 
with $z_{\rm min}$ and $S_{c}=0$, 
\cite{OH99} found a free-free distortion 
$\Delta T_{ff}= 3.4 \times 10^{-3}$~K at 2~GHz, corresponding to 
a free-free distortion parameter $y_{B}\simeq 1.5\times 10^{-6}$,
well within the observational capability of DIMES \cite{KOG96,BS03a}.
 
\section{CONCLUSION}

Although not specifically devoted to CMB studies, because of its 
high resolution and the limited high frequency coverage, 
the extreme sensitivity and resolution of SKA may be fruitfully 
used for a detailed mapping of the 
thermal plasma properties in the 
intergalactic and intracluster medium and at
galaxy scale.

On the basis of the existing literature, we discussed 
the SKA contribution to the understanding
of some topics of relevant interest in cosmology, from 
the detailed mapping of the SZ effect towards cluster of galaxies,
to the number counts and imaging possibilities 
of the SZ effect from early quasars and proto-galactic gas 
and of the free-free emission from early ionized halos.
We also discussed the contribution of SKA 
to the future CMB spectrum space experiments, devoted 
to the comprehension
of the plasma thermal history at early epochs,
through the extremely accurate control of extragalactic radio source
counts at very faint fluxes.

\appendix
\section{SKA SENSITIVITY IN TERMS OF ANGULAR POWER SPECTRUM}
\label{skaperformance}

The properties of a large population of astrophysical objects 
can be studied by exploiting the details of the sources in dedicated 
fields and by analyzing the field statistical properties.

The statistics of temperature anisotropy is typically analyzed in 
spherical harmonics $Y_{\ell \, m}$:
                                                                                
\begin{equation}
\frac{\delta T}{T}(\hat{\gamma})=\sum_{\ell = 1}^{\infty}
\sum_{m= -\ell}^{\ell} a_{\ell \, m}
Y_{\ell \, m}(\hat{\gamma}) \, ,
\end{equation}
where $\hat \gamma$ is the observation unit direction vector.
By assuming isotropy around the observer, $a_{\ell m}$ should
have zero mean, $\langle a_{\ell m} \rangle = 0$, 
and variance $C_\ell$~\footnote{
Roughly speaking, a multipole $\sim \ell$ corresponds to an angular
scale $\vartheta / {\rm deg} = 180 / \ell$.
In this convention $C_\ell$ is dimensionless. 
It is also useful to bear in mind that the same 
expansion in spherical harmonics can be considered 
for the (physical) temperature fluctuation 
$\delta T$, instead of the dimensionless 
temperature fluctuation $\delta T /T$; in this case
the physical dimension of $C_\ell$ will be the square of 
a temperature. 
Given the CMB monopole temperature 
of 2.725~K \cite{mather99}, the dimensionless 
$C_\ell$ 
will be $\simeq 7.4\times10^{12}$ smaller than the
$C_\ell$ expressed in terms of $\mu$K$^2$ (in thermodynamic temperature). 
} given by:
\begin{equation}
C_\ell \equiv\langle\mid a_{\ell m}^2\mid\rangle
= \frac{1}{2 \ell + 1}\sum_m a_{\ell m}^2 \, .
\end{equation}

The correlation function of the temperature anisotropy,
$C(\theta) \equiv 
\langle\frac{\delta T}{T}(\hat{\gamma_1})
\frac{\delta T}{T}(\hat{\gamma_2})\rangle$, 
is related to the angular power spectrum by the equation: 
\begin{equation}
C(\theta) 
= \frac{1}{4\pi}\sum_{l}(2l+1) \, C_\ell \, P_\ell (\cos \theta) \, ;
\end{equation}
here $\cos \theta = \hat{\gamma_1} \cdot \hat{\gamma_2}$
and $P_\ell$ is the Legendre
polynomial.

\begin{figure}[htb]
\includegraphics[angle=0.,width=7.cm,height=7.cm]{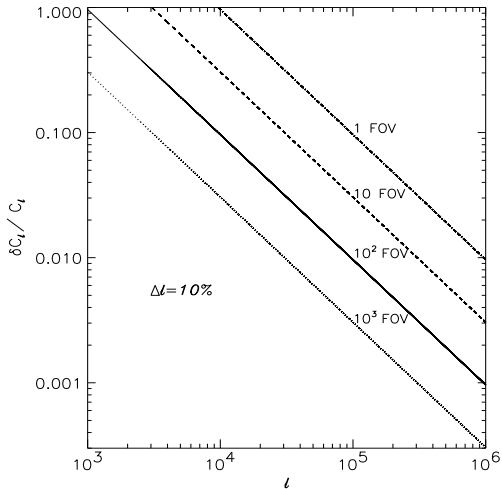}
\caption{Relative uncertainty on CMB angular power spectrum 
recovery from combined cosmic and sampling variance 
for some representative sky coverages.}
\label{fig:cosm_samp_var}
\end{figure}

At SKA resolution and sensitivity the standard approaches to 
compute the angular power spectrum of anisotropies from classes 
of astrophysical sources, typically based on the assumption that 
sources above the detection level do not significantly merge in 
the same image pixel, are no longer valid since the expected 
number of (no longer point-like) sources per resolution element may be 
large and the source diffuse emission extends over several (or many) 
image pixels (see, e.g., Fig.~\ref{fig:n_sz_gal}). 
Further studies are then necessary to characterize 
the correlation properties of source populations as 
observed at SKA resolution and sensitivity. 

In spite of this uncertainty, it is important to characterize
the nominal SKA capability to recover these kinds of statistical 
information.   

We then ``rewrite'' here the typical SKA sensitivity 
in terms of accuracy in the recovery of the angular power spectrum.

Since each given anisotropy field is a single
realization of a stochastic process, 
it may be different from the average over the ensemble
of all possible realizations of the given (true) 
model with given parameters.
This translates into the fact that the $a_{\ell m}$ coefficients
are random variables (possibly following a Gaussian distribution), at a
given $\ell$, and therefore
their variance, $C_\ell$, is $\chi^2$ distributed with $2\ell +1$ degrees 
of freedom. The relative variance $\delta C_\ell$ on $C_\ell$ is equal to
$\sqrt{2/(2\ell+1)}$ which is quite relevant at low $\ell$ because of the
relatively small number of available modes.
This is the so-called ``cosmic variance'' that defines the ultimate limit
on the accuracy at which a given 
model defined by
an appropriate set of parameters can be constrained
by the 
angular power spectrum.
Another similar variance in 
anisotropy experiments
is related to the sky coverage since the detailed 
anisotropy
statistical properties may depend on the considered sky patch.
This variance depends on the observed sky fraction, $f_{\rm sky}$.
At the largest multipoles achievable with a given experiment 
the most relevant uncertainties are related to the experiment resolution and
sensitivity. All these terms contribute to the final uncertainty on the
angular power spectrum according to \cite{knox95}:
\begin{equation}
\frac{\delta C_\ell}{C_\ell} = \sqrt{\frac{2}{f_{\rm sky}(2\ell+1)}}\left[
1+\frac{A\sigma^2}{NC_\ell W_\ell}\right]\, ,
\label{fullvariance}
\end{equation}
where $A$ is the size of the surveyed area, $\sigma$ is the rms noise per 
pixel, $N$ is the total number of observed pixel, and $W_\ell$ is the beam 
window function. For a symmetric Gaussian beam
$W_\ell = {\rm exp}(-\ell(\ell+1)\sigma_{\rm B}^2)$ where
$\sigma_{\rm B} = {\rm FWHM}/\sqrt{8{\rm ln}2}$
defines the beam resolution.
In the limit of an experiment with infinite sensitivity
($\sigma=0$) $\delta C_\ell /C_\ell$ is determined only by cosmic 
and sampling variance; in this limit, by assuming also full sky 
coverage ($f_{\rm sky}=1$), $\delta C_\ell /C_\ell$ is determined only by 
the cosmic variance.
  
At $\simeq 20$~GHz a typical SKA field of view (FOV) is of $\simeq 4'
\times 4'$ and a sky coverage of $\simeq 10^2$
FOV corresponds to 
about $\simeq 40' \times 40'$, possibly achievable 
with mosaicing techniques.

\begin{figure}[htb]
\includegraphics[angle=0.,width=7.cm,height=7.cm]{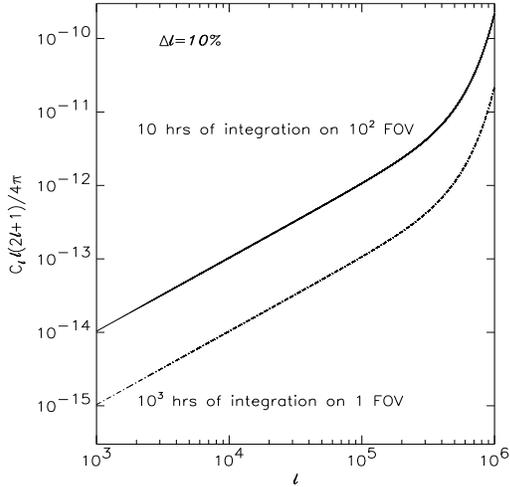}
\caption{Absolute uncertainty on dimensionless CMB angular power spectrum 
recovery due to the instrumental noise 
for some reference cases (see also the text).}
\label{fig:instrum_sens_adim}
\end{figure}

\begin{figure}[htb]
\includegraphics[angle=0.,width=7.cm,height=7.cm]{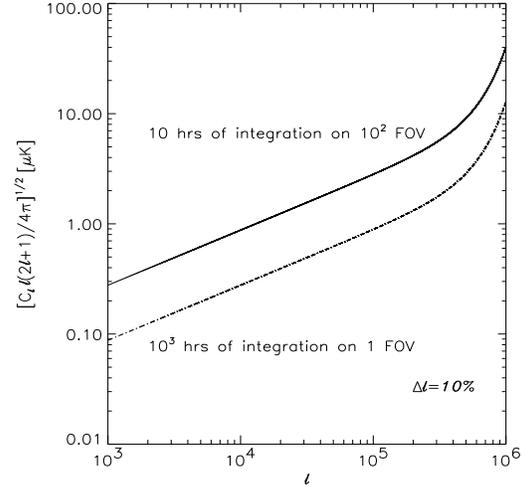}
\caption{The same as in Fig.~\ref{fig:instrum_sens_adim},
but in terms of the temperature fluctuation 
$\sqrt{C_\ell \ell (2 \ell+1)/4\pi}$.}
\label{fig:instrum_sens}
\end{figure}

The relative error $\delta C_\ell /C_\ell$ due to
cosmic and sampling variance is shown in Fig.~\ref{fig:cosm_samp_var}
by considering a binning of $\simeq 10$\% in $\ell$ which 
allows to reduce $\delta C_\ell /C_\ell$ by a factor 
$\sqrt{0.1 \ell}$ and is not 
critical, being quite smooth the $C_\ell$ dependence on $\ell$ 
for the kinds of anisotropies considered in this work.
As evident the sky coverage corresponding to 
$\sim 10^2$ SKA FOV 
(or to $\sim 1$ SKA FOV) 
allows to reach a relative ``fundamental'' uncertainty less than $\simeq 10$\%
at $\ell \simeq 10^4$ (or at $\simeq 10^5$) and significantly smaller at 
larger multipoles.

The smallest angular scales (or the highest multipoles) 
achievable with a good accuracy are set by
the instrument resolution and sensitivity.

By using about 50\% of the collecting area~\footnote{In this context, 
the whole SKA collecting area could represent 
a real improvement only provided that surface brightness (or temperature)  
anisotropy maps at higher resolution could be degraded 
at lower resolution to improve the rms (white noise) sensitivity 
per resolution element (according to the ratio of the corresponding
FWHM) by averaging the brightness temperatures of the small pixels
which constitute a larger pixel, a point that should be accurately 
verified through detailed simulations.}
SKA would allow the mapping of diffuse anisotropies 
up to $\ell \sim 10^6$.
We will use this case as a reference.

The absolute error $\delta C_\ell$ due to
the instrumental noise is shown in 
Figs.~\ref{fig:instrum_sens_adim} and \ref{fig:instrum_sens}, 
again
by considering a binning of $\simeq 10$\% in $\ell$,
respectively 
in terms of 
$C_\ell \ell (2 \ell+1)/4\pi$ with dimensionless $C_\ell$
and of $\sqrt{C_\ell \ell (2 \ell+1)/4\pi}$ 
with $C_\ell$ in $\mu$K$^2$
(note that we do not refer here to the noise angular power spectrum but 
to the uncertainty on the signal angular power spectrum recovery, 
i.e. to the rms of ``residual'' noise angular power spectrum
after the subtraction of its expectation value obtained 
on the basis of the accurate knowledge of the instrumental noise
properties, as necessary in CMB anisotropy experiments).

\vskip 1cm

\noindent
{\bf ACKNOWLEDGMENTS} It is a pleasure to thank 
A.~Cavaliere, L.~Danese, C.~Fanti, F.~Finelli, G.L.~Granato, A.~Lapi, 
N.~Mandolesi, P.~Platania, L.A.~Popa, R.~Salvaterra, 
and L.~Silva for collaboration and discussions.
The use of the {\sc cmbfast} code is acknowledged.

\end{document}